\documentclass[paper]{elsarticle}

\usepackage{lineno,hyperref}
\usepackage{amsmath}
\usepackage{amsfonts}
\usepackage{amssymb}
\usepackage{makeidx}
\usepackage{graphicx}
\usepackage{mathrsfs}
\usepackage{algorithm}
\usepackage{algorithmic}
\usepackage{booktabs}
\usepackage{accents}
\usepackage{siunitx}
\usepackage{gensymb}
\usepackage{graphicx}
\usepackage{subcaption}
\usepackage{xcolor}
\usepackage{chngcntr}

\modulolinenumbers[5]
\newcommand\munderbar[1]{%
	\underaccent{\bar}{#1}}

\journal{Journal of Process Control}
\let\linenumbers\nolinenumbers\nolinenumbers
\begin{document}

\setcounter{secnumdepth}{5}
\begin{frontmatter}

	\title{A Hybrid Gaussian Process Approach to Robust Economic Model Predictive Control}
	%\tnotetext[mytitlenote]{Fully documented templates are available in the elsarticle package on \href{http://www.ctan.org/tex-archive/macros/latex/contrib/elsarticle}{CTAN}.}

	%% Group authors per affiliation:
	%\author{R, N\fnref{myfootnote}}
	%\address{vancouver, UBC}
	%\fntext[myfootnote]{sth}

	%% or include affiliations in footnotes:
	\author[mymainaddress]{Mohammadreza Rostam\corref{mycorrespondingauthor}}
	\cortext[mycorrespondingauthor]{Corresponding author}
	\ead{reza.rostam@mech.ubc.ca}
	%\ead[url]{www.elsevier.com}
	\author[mymainaddress]{Ryozo Nagamune}
	\address[mymainaddress]{Department of Mechanical Engineering, The University of British Columbia, Vancouver, BC V6T 1Z4, Canada}
	\author[mymainaddress2]{Vladimir Grebenyuk}
	\address[mymainaddress2]{Ascent Systems Technologies, Heffley Creek, BC V0E 1Z0, Canada}

	\begin{abstract}
		This paper proposes a hybrid Gaussian process (GP) approach to robust economic model predictive control under unknown future disturbances in order to reduce the conservatism of the controller. The proposed hybrid GP is a combination of two well-known methods, namely, kernel composition and nonlinear auto-regressive. A switching mechanism is employed to select one of these methods for disturbance prediction after analyzing the prediction outcomes. The hybrid GP is intended to detect not only patterns but also unexpected behaviors in the unknown disturbances by using past disturbance measurements. \textcolor{black}{A novel forgetting factor concept is also utilized in the hybrid GP, giving less weight to older measurements, in order to increase prediction accuracy based on recent disturbances values.} The detected disturbance information is used to reduce prediction uncertainty in economic model predictive controllers systematically. The simulation results show that the proposed method can improve the overall performance of an economic model predictive controller compared to other GP-based methods in cases when disturbances have discernible patterns.
	\end{abstract}


\end{frontmatter}

\linenumbers

\section{Introduction}
Economic Model Predictive Control (EMPC) is becoming more and more popular in industrial applications due to its ability to achieve both the optimal output targets and optimal control inputs simultaneously, by respecting constraints on state variables~\cite{Ellis2017}. The performance of EMPC has been evaluated for different plants such as chemical process systems~\cite{Santander2016, Zeng2015} and building energy systems~\cite{Ma2012, Serale2018}. Regardless of applications, the performances of EMPC depends crucially on the accuracy of the prediction step. Unknown future disturbances pose a challenge when predicting future state trajectories precisely since, in general, they are not accurately described by any model. In fact, it was shown that EMPC may perform worse than a simple proportional-integral-derivative (PID) controller in terms of both energy savings and constraint violations once actual disturbances differ from predicted ones~\cite{MaFeb}.

To tackle the challenge of unknown future disturbances in EMPC, several methods for improving the disturbance prediction have been proposed so far, including auto-regressive integrated moving average models~\cite{Lin2005, Box2015} and artificial neural networks~\cite{Doganis2008}. A downside to most of these methods is the fact that they cannot take the uncertainty of prediction into account. Thus, in these methods, there is no systematic way to obtain a reasonable uncertainty range for disturbances which is useful information to alleviate the conservatism in EMPC.

Gaussian Process (GP) is a probabilistic and non-parametric method that takes into consideration the uncertainty of a process systematically~\cite{Rasmussen2006}. This feature makes GP one of the best options for forecasting unpredictable stochastic variables. GP has been employed for wind speed forecasting for wind power generation~\cite{Mori_2008,Chen_2014, Yan_2016, Zhang_2016}, short-term load forecasting~\cite{Mori}, and energy savings estimation~\cite{Heo2012}. In these works, GP outperformed the existing approaches such as artificial neural networks for short-term predictions. In the case of long-term predictions, however, a standard GP cannot properly handle the uncertainty at future times far from the prediction moment~\cite{Yan_2009}.

For long-term prediction, a simple approach is to employ one-step-ahead predictions successively, where the outputs of each step act as the inputs for the next step of prediction. This method is called Na\"ive multiple-step-ahead prediction, and due to the accumulation of uncertainties, it is not suitable for long-term predictions, giving an unacceptably large uncertainty in most cases~\cite{Girard2003}.

The \textit{Nonlinear Auto-regressive (NAR) method}~\cite{Girard2003} is another GP-based method, in which there is one unique GP model for each step-ahead prediction~\cite{Frigola2015}. Hachino~et~al.~experimentally demonstrated that the NAR method is accurate even in the presence of measurement noise~\cite{Hachino2015}. Wang~et~al.~employed the uncertainty of the state prediction given by a model based on the NAR method to determine the worst-case scenario and make the model predictive controller robust~\cite{Wang165,Wang2018}. However, it was pointed out in~\cite{Girard2003} that the NAR method ignores the accumulating prediction variance, thereby leading to an unduly small uncertainty attached to the forecast.

Recently, another method for long-term predictions was introduced, called the \textit{Kernel Composition (KC) method}~\cite{Brahim-Belhouari2004, Duvenaud2013}. The main idea behind the KC method is to use a combination of kernels in order to capture the patterns of disturbances~\cite{Rasmussen2006}. Maritz~et~al.~adopted the KC method to quantify the uncertainties that govern an energy system, and showed the robustness of this method for energy data regressive~\cite{Maritz2018}. The KC method demonstrated an excellent prediction in the presence of noticeable patterns in time-series data~\cite{Duvenaud2014}. Nevertheless, the GP model in the KC method has extra complexity due to the increased number of hyperparameters, which makes the training step prone to failure due to the non-convex optimization problem. Thus, the KC method is inappropriate for real-time applications in general.

In this paper, to overcome the drawbacks of the unrealistic uncertainty ranges in the NAR method and the risk of training failure in the KC method, a novel hybrid GP method is proposed in which the NAR and the KC methods are combined using a switching strategy. In this strategy, the KC method is employed first to predict future disturbances in order to avoid the unnecessarily large uncertainty range of the NAR method. Then, the prediction outcome is assessed to see whether the prediction seems rational considering previous disturbances. Rationality criteria include the accuracy of  disturbance pattern detection and unreasonable growth of the uncertainty range of the prediction. In the case of an irrational KC result, the NAR method is activated instead. Therefore, not only does the proposed predictor take advantage of the patterns that exist in disturbances, it also identifies failed-in-training models online and avoids using these failed models for forecasting. \textcolor{black}{In addition, to enhance the performance of the internal KC method when there are two separate patterns in the past measured disturbances, a forgetting factor term is added to the KC method so that the recent measured data points have more weight for predicting the future disturbances.}

The contribution of this paper is three-fold. First, it introduces a structure of a disturbance predictor which combines both the NAR and KC methods. Second, it proposes a switching rule in order to utilize the best prediction at each time instant. \textcolor{black}{Last, it integrates the forgetting factor concept and the internal KC method to adapt more effectively to with changes in the disturbance pattern through the training horizon.}

This paper is organized as follows. In Section~\ref{sec:problem}, the problem of predicting the disturbances in robust EMPC formulation is stated. In Section~\ref{sec:REMPC}, a hybrid method to tackle the disturbance prediction problem is proposed and described in detail. The simulation results of applying the proposed hybrid GP-based robust EMPC to a tank-heater system are presented and discussed in Section~\ref{sec:case}.

The notations in this paper are as follows. $\mathbb{R}^{n \times m}$ is the set of real matrices of size $n\times m$ and$\mathbb{R}_{+}$ is the set of positive real numbers. For a matrix $\textbf{M}$, $[\textbf{M}]_{i,j}$ means the $(i,j)$-element of $\textbf{M}$, and for a vector $\mathbf{v}$, $\mathbf{v}_i$ means the $i$-th element of~$\mathbf{v}$.
\section{Problem statement} \label{sec:problem}
Consider the general class of discrete-time nonlinear time-invariant systems represented by
\begin{equation}
	x_{k+1} = f(x_k, u_k, w_k),
\end{equation}
where $x_k \in\mathbb{R}^{n}$, $u_k \in\mathbb{R}^{n_u}$, and $w_k \in\mathbb{R}$ \footnote{Without loss of generality, it is assumed that there is just one disturbance here. In the case of multiple disturbances, one may extend the proposed method by creating a GP model for each of the disturbances independently.} are the state vector, the decision vector, and the disturbance value at the time instance $k$, and $f$ is a nonlinear function with respect to these vectors. Given the above system, one can define a Robust Economic Model Predictive Control (REMPC) problem at each time instant $k$ with the prediction horizon $N_p$ as follows \cite{Bemporad}:
%Each EMPC has a performance index, $l_e$, and manages the system in a manner to obtain the best performance index based on a prediction. The proposed Robust EMPC strategy is implemented solving the following optimization problem:
\begin{equation} \label{eq:obj}
	\begin{array}{l}
		\displaystyle\min_{
			\tiny \begin{array}{c}
				\{ u_{k+i|k} \in \mathcal{U}\}_{i=1}^{N_p}, \\
				\{ (\munderbar \gamma_{i}, \bar \gamma_{i})\}_{i=1}^{N_p}
			\end{array} }\max_{\{ w_{k+i|k} \in \mathcal{W}_{k+i}\}_{i=1}^{N_p}} \sum_{i=1}^{N_p} \mathbb{J}_i(x_{k+i|k}, u_{k+i|k}, \munderbar \gamma_{i}, \bar \gamma_{i}) \\%\mathbb{J}(\mathbf X, \mathbf U, \mathbf W, \mathbf \Gamma) \\
		\hspace{7em}\mbox{subject to }\left\{
		\begin{array}{l}
			x_{k+i+1|k} = f(x_{k+i|k}, u_{k+i|k}, w_{k+i|k}) \\
			\munderbar{x} - \munderbar \gamma_{i} \leq x_{k+i|k} \leq \bar{x} + \bar \gamma_{i},  i = 1, \ldots, N_p.
		\end{array}
		\right.
	\end{array}
\end{equation}
% \begin{align} \label{eq:obj2}
% 		\displaystyle\min_{
% 			\tiny \begin{array}{c}
% 				\{ u_{k+i} \}_{i=1}^{N_p} \subseteq \mathcal{U}, \\
% 				\{ (\munderbar \gamma_{i}, \bar \gamma_{i})\}_{i=1}^{N_p}
% 			\end{array} }\max_{\{ w_{k+i} \in \mathcal{W}_{k+i-1}\}_{i=1}^{N_p}} & \qquad \mathbb{J}(\mathbf X, \mathbf U, \mathbf W, \mathbf \Gamma) \\
% 		\mbox{subject to } & \left\{
% 		\begin{array}{l}
% 			x_{k+i+1|k} = f_d(x_{k+i|k}, u_{k+i|k}, w_{k+i|k}) \\
% 			\munderbar{x} - \munderbar \gamma_{i} \leq x_{k+i|k} \leq \bar{x} + \bar \gamma_{i},  i = 1, \ldots, N_p,
% 		\end{array}
% 		\right.
% \end{align}
The notation $v_{k+i|k}$ indicates the value of the vector variable $v$ at the instant $k+i$ calculated at instant $k$. Due to the physical limitations of actuators, decision variables have to be chosen in the range of $\mathcal{U}:=\left[ \munderbar{u}, \bar u \right]$, where $\munderbar{u}\in\mathbb{R}^{n_u}$ and $\bar u\in\mathbb{R}^{n_u}$ are the lower and upper limits of the decision variables. As for the state variables, to have a value between $\munderbar{x}$ and $\bar{x}$, soft constraints are applied here by using the slack vector variables $\bar \gamma_{i} \in \mathbb{R}^{n}_{+}$ and $\munderbar \gamma_{i} \in \mathbb{R}^{n}_{+}$.
% As for the state variables, forcing the optimization problem to keep the states within the  range, $\left[ \munderbar{x}, \bar x \right]$, by introducing hard constraints on the states might lead to an infeasible optimization problem. Therefore, soft constraints 
%, and these values are passed sequentially to the objective function in a matrix form, $\mathbf \Gamma$.

The objective function in~(\ref{eq:obj}) is divided into two separate terms in the following form:%with an additional penalty term for the state constraints.
\begin{align}
	\mathbb{J}_i(x,u,\munderbar{\gamma},\bar{\gamma}) \triangleq \mathbb{J}^{EC}_{i}(x,u)+\mathbb{J}^{CV}_{i}(\munderbar{\gamma},\bar{\gamma})
\end{align}
Here, the first term, $\mathbb{J}_{e,i}$, corresponds to the economical cost function of the process. As an example, a quadratic form of the decision variables can be considered for this term in the case of EMPC;
\begin{equation}
	\mathbb{J}^{\mathrm{EC}}_{i}(x,u) \triangleq u^T \textbf{R}_i u,
\end{equation}
where $\textbf{R}_i\in\mathbb{R}^{n_u\times n_u}$ is a positive definite matrix. The second term, $\mathbb{J}_{CV,i}$, indicates the state constraint violations, and it is defined as
% \begin{align}
% 	\mathbb{J}(\mathbf X, \mathbf U, \mathbf W) \triangleq \mathbb{J}_e + \mathbb{J}_{CV}\sum_{i=1}^{N_p} \left[ \mathbb{J}_e(x_{k+i|k}, u_{k+i|k}, w_{k+i|k}) \right] + CV(N_p), \label{eq:obj}
% \end{align}
% where $CV$ is the constraint violation index and it is defined as follows.
\begin{equation} \label{eq:CV}
	\mathbb{J}^{\mathrm{CV}}_{i}(\munderbar{\gamma},\bar{\gamma}) \triangleq \munderbar \eta_{i} \munderbar \gamma + \bar \eta_{i} \bar \gamma,
\end{equation}
where $\munderbar \eta_{i} \in\mathbb{R}^{1 \times N_p}$ and $\bar \eta_{i} \in\mathbb{R}^{1 \times N_p}$ are the positive-element vectors to penalize the violation of the state constraints.

The REMPC formulation in~(\ref{eq:obj}) has nothing new; in fact, there are quite a few papers on how to deal with the min-max problem described in~(\ref{eq:obj}) \cite{Loefberg2003,de_la_Pe_a_2005,Zeman}. However, most of them solved the problem given a fixed range for the values of the disturbances, $\mathcal{W}= \left[  \munderbar{w}, \bar w \right]$ where $\underbar{w}$ and $\bar{w}$ are the uniform lower and upper bounds, respectively. This assumption can lead to conservative controllers if the range of the disturbance is changing with time or state. Since it is of practical importance to reduce the conservatism in REMPC, identifying an optimal disturbance prediction range from time $k+1$ to $k+N_p$ at time instant $k$, denoted by $\mathbb{W}_k\triangleq\{\mathcal{W}_{k+i}\}_{i=1}^{N_p}$, which we call an ``envelope'' hereafter, for the entire prediction horizon is essential.

\begin{figure}[h]
	\centering
	\begin{subfigure}{0.6\textwidth}
		\includegraphics[width=\textwidth]{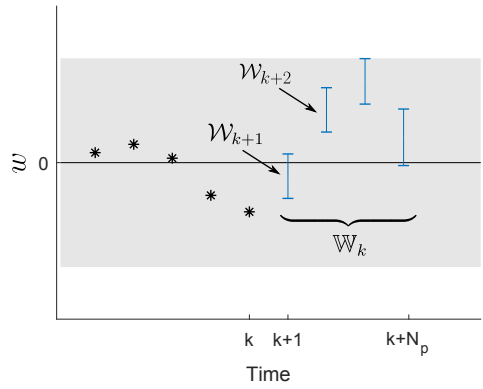}
		\caption{Uncertainty envelope, $\mathbb{W}_k$, evaluated at time $k$ (blue error bars)}
		\label{fig:explain1}
	\end{subfigure}
	\begin{subfigure}{0.6\textwidth}
		\includegraphics[width=\textwidth]{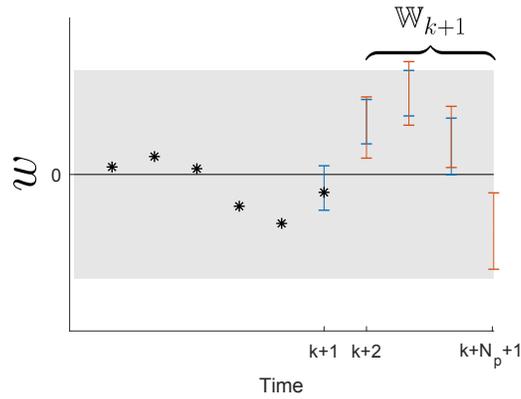}
		\caption{Uncertainty envelope, $\mathbb{W}_{k+1}$, evaluated at time $k+1$ (red error bars)}
		\label{fig:explain2}
	\end{subfigure}
	\caption{Two successive estimations of the future disturbances envelope (the grey shaded areas correspond to a fixed range disturbance)}
	\label{fig:explain}
\end{figure}

To give a pictorial explanation of non-convservative ranges $\mathcal{W}_{k+i}$, let us look at Fig.~\ref{fig:explain}. At time $k$, given the past and present disturbance measurements, we seek for an envelope for the future disturbances at time $k$, indicated by the blue error bars in Fig.~\ref{fig:explain1}. Note that for the sake of comparison, a potential fixed range disturbance is also drawn by the grey shaded area, to illustrate how conservative the fixed-range assumption could be. At the next time instant, $k+1$, the new ``present'' disturbance, $w_{k+1}$, is measured, and a brand-new envelope, $\mathbb{W}_{k+1}$, can be defined for the new prediction horizon as illustrated with red color in Fig.~\ref{fig:explain2}.

In the next section, a novel approach is proposed in order to determine $\mathbb{W}_{k}$ optimally at each time instant $k$ in a real-time fashion.

% \begin{figure}
% 	\ContinuedFloat
% 	\centering
% 	~ %add desired spacing between images, e. g. ~, \quad, \qquad, \hfill etc.
% 	%(or a blank line to force the subfigure onto a new line)
% 	\begin{subfigure}{0.7\textwidth}
% 		\includegraphics[width=\textwidth]{figures/explain_k+1}
% 		\caption{Uncertainty envelope, $\mathcal{W}_{k+1}$, evaluated at time $k+1$ (red shaded area)}
% 		\label{fig:explain2}
% 	\end{subfigure}
% 	\caption{Two successive estimations of the future disturbances envelope (the grey shaded areas correspond to a fixed range disturbance)}
% 	\label{fig:explain}
% \end{figure}

~ %add desired spacing between images, e. g. ~, \quad, \qquad, \hfill etc.
%(or a blank line to force the subfigure onto a new line)
% To solve , a range for the values of disturbances, $\mathcal{W}$, is required.  
% It means that at first, it is required to find the maximum of the objective function with respect to the disturbance vector.  In order to obtain the worst-case scenario, a rational range for the disturbances should be identified in advance.
% The problem of robust model predictive control is solved is several paper. Despite of it, all paper consider that a set of W is given. If we consider a big range of W, the results leads to a conservative control which can use a lot of energy. This situation is not desireable especially in EMPC case.
%A popular method for handling state and output constraints in a model predictive control (MPC) algorithm is to use "soft constraints", in which  This this manner, we can avoid infeasibility of the optimization problem.
%\begin{align}
%& \underset{\textbf{u}}{\text{min}}
%& & \max J \triangleq \sum_{i=1}^{N_p} \left[ l_e(x_{k+i|k}, u_{k+i|k}, w_{k+i|k}) + c_{n,1}\gamma_{n,1} + c_{n,2}\gamma_{n,2} \right]\\
%& \text{subject to}
%& & \munderbar{u} \leq u_{k+i|k} \leq \bar u. \\
%&
%%& & w_{k+i|k} \sim \mathcal{N}(m_{k+i-1|k}, c_{k+i-1|k}). \\
%%&
%& & \gamma_{i,1} = \max \{ x_{k+i|k} - \bar{x}, 0 \}  \\
%&
%& & \gamma_{i,2} = \max \{ \munderbar{x} - x_{k+i|k}, 0 \}
%\end{align}
\section{REMPC based on GP} \label{sec:REMPC}
Gaussian process approaches assign a Gaussian (normal) distribution, instead of a value, to each data point throughout the prediction horizon. Hence, a reasonable ranges for $\mathcal{W}_{k+i}$ can be obtained by specifying a confidence level for the Gaussian distribution. This characteristic makes GP-based approaches suitable to determine the uncertainty range of disturbances in REMPC.

\subsection{Proposed hybrid GP structure for REMPC}
\begin{figure}
	\centering
	\includegraphics[width=0.6\linewidth]{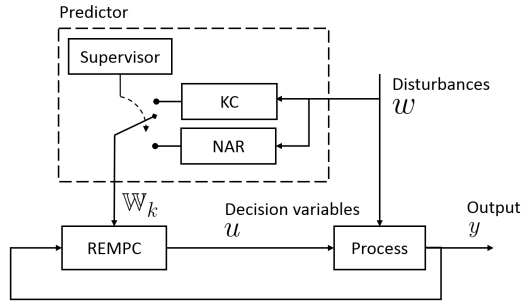}
	\caption{Block diagram of a robust economic model predictive controller along with a predictor using the hybrid method}
	\label{fig:diagram2}
\end{figure}

The proposed predictor for the set $\mathbb{W}_k$ of the uncertainty range $\mathcal{W}_{k+i}$ consists of two GP-based methods called the kernel composition~(KC) and the nonlinear auto-regressive~(NAR), and a supervisory block. As shown in Fig.~\ref{fig:diagram2}, the present disturbances are assumed to be measurable by sensors and input to the predictor. In the case of unmeasurable disturbances, one may utilize an observer block in order to evaluate the values of disturbances~\cite{Aldeen2008}. Since both the KC and the NAR methods produce probabilistic models, the output of these two blocks describe two distinct envelopes for future disturbances. Next, in order to choose one of the obtained envelopes, the supervisory block generates a switching signal by following a switching rule. If the KC method predicts satisfactorily according to the switching rule, the REMPC uses its output. Otherwise, the REMPC will switch to the output of the NAR method. Therefore, the switching rule must be capable of recognizing whether a forecast is valid considering the past and present disturbances.

In order to understand how the predictor works in detail, we first introduce some preliminary in Section~3.2, and using the preliminary concepts, we explain the KC-block and NAR-block in Section~3.3 and the Supervisor-block in Section~3.4.

\subsection{Probabilistic GP model}
The disturbance is considered to be a general function,
\begin{equation} \label{eq:disturbance}
	w(t)|_{t=kT} = w_k + \epsilon_k,
\end{equation}
where $w(t)$ and $w_k$ are the actual disturbance and the measured disturbance at time $t=kT$, respectively. In~(\ref{eq:disturbance}), $T$ is the sampling period, $k$ is the discrete time index, and $\epsilon_k$ is additive independent and identically distributed Gaussian noise with variance denoted by $\sigma$,
\begin{equation}
	\epsilon_k \sim \mathcal{N}(0, \sigma^2).
\end{equation}
The form in~(\ref{eq:disturbance}) is used to take noisy observations into account and it is assumed that the variance of the noise is constant through a prediction horizon. A Gaussian process is a collection of random variables, any finite number of which have a joint Gaussian distribution. Thus, the probability of both training points and prediction points at time $k$ can be written as follows \cite{Rasmussen2006}:
\begin{equation}\label{eq:Joint}
	p (\begin{bmatrix} w(\textbf{t}_\text{t}) \\
		w(\textbf{t}_\text{p})\end{bmatrix} ) = \mathcal{N}\Biggl(\begin{bmatrix} {\mu}(\textbf{t}_\text{t}) \\
			{\mu}(\textbf{t}_\text{p})\end{bmatrix} ,\begin{bmatrix} \textbf{K}_\hbar(\textbf{t}_\text{t},\textbf{t}_\text{t}) + \sigma^2 \textbf{I} & \textbf{K}_\hbar^T(\textbf{t}_\text{p},\textbf{t}_\text{t}) \\
			\textbf{K}_\hbar(\textbf{t}_\text{p},\textbf{t}_\text{t})                       & \textbf{K}_\hbar(\textbf{t}_\text{p},\textbf{t}_\text{p})\end{bmatrix}\Biggr),
\end{equation}
where $\textbf{t}_\text{t}$ and $\textbf{t}_\text{p}$ are the past time vector and the future time vector, respectively:
\begin{align}
	\textbf{t}_\text{t} & \triangleq \left[ (k-N_t)T, (k-N_t+1)T, \cdots, kT\right], \\
	\textbf{t}_\text{p} & \triangleq \left[ (k+1)T, \cdots, (k+N_p)T \right],
\end{align}
%Can you use better notations? for example $\textbf{t}_{past}, \textbf{t}_{future}$ $f(\textbf{t})$ is the vector of training points and $\textbf{t}_*$ is the prediction time vector and 
with the training horizon length $N_t$ and the prediction horizon length $N_p$. This training horizon length, $N_t$, has a significant influence on the estimation results. As the training horizon increases, there will be more data available for training, which improves the pattern finding process. If, however, a sudden pattern change occurs in the disturbance, it takes more time for the predictor to adjust itself to the new pattern. In~(\ref{eq:Joint}), $\mu$ is the a priori mean value of the Gaussian distribution. $\textbf{K}_\hbar$ is the covariance matrix, also known as the kernel since it measures a degree of similarity between points in the dataset. Each kernel has some tuning parameters which are called \textit{hyperparameters} and are denoted by $\hbar$. Kernels express our initial idea over the function that we want to estimate. Any function could be a covariance function as long as the resulting covariance matrix is positive semi-definite. The most common kernel is the radial basis function (RBF) given by
\begin{equation} \label{eq:RBFkernel}
	\left[ \textbf{K}_{\hbar=[\delta, \lambda]} ^ {\text{RBF}} \right]_{i,j} (\textbf{t}, \textbf{t}') \triangleq \delta^2 \exp \left( -\frac{1}{2 \lambda^{2}} (\textbf{t}_i-\textbf{t}'_j)^2 \right).
\end{equation}
To predict the values of future disturbances, given the past and present measured disturbances, a conditional probability distribution is required and it is derived from~(\ref{eq:Joint}) \cite{Murphy2012}:
\begin{equation} \label{eq:NGP}
	p(w(\textbf{t}_\text{p})|w(\textbf{t}_\text{f}),\hbar) = \mathcal{N}(\textbf{m}_\text{p}(\textbf{t}_\text{p}), \textbf{C}_\text{p}(\textbf{t}_\text{p})),
\end{equation}
where
\begin{subequations} \label{eq:meancov}
	\begin{align}
		\textbf{m}_\text{p}(\textbf{t}_\text{p}) & \triangleq {\mu} (\textbf{t}_\text{p}) + \textbf{K}_\hbar(\textbf{t}_\text{p},\textbf{t}_\text{t}) \textbf{K}_\hbar(\textbf{t}_\text{t},\textbf{t}_\text{t})^{-1} \left[ w(\textbf{t}_\text{t}) - {\mu}(\textbf{t}_\text{t})\right],                           \\
		\textbf{C}_\text{p}(\textbf{t}_\text{p}) & \triangleq \textbf{K}_\hbar(\textbf{t}_\text{p},\textbf{t}_\text{p}) - \textbf{K}_\hbar(\textbf{t}_\text{p},\textbf{t}_\text{t}) \textbf{K}_\hbar(\textbf{t}_\text{t},\textbf{t}_\text{t})^{-1} \textbf{K}_\hbar(\textbf{t}_\text{p},\textbf{t}_\text{t})^{T}.
	\end{align}
\end{subequations}
For the sake of simplicity, hereafter, a Guassian process model is denoted as
\begin{equation}
	w(\textbf{t}_\text{p}) \sim \mathcal{GP}(\textbf{m}_\text{p}(\textbf{t}_\text{p}), \textbf{C}_\text{p}(\textbf{t}_\text{p})).
\end{equation}
%Since these two functions together indicate a Gaussian distribution from which an envelope is derived using a confidence level. Therefore, 
%It means the function $f(t)$ is drawn from a Gaussian process with some probability.
\subsubsection{Disturbance uncertainty range envelope}
As mentioned before, to extract a range from a given Gaussian distribution, the confidence level ($\beta$) concept is utilized. The higher confidence level is for a disturbance $w(t)$ to lie within a range, the lower the risk ($\alpha$) of the data that appears outside of the range. This can be mathematically expressed as

\begin{equation}\label{eq:prisk}
	p(\munderbar w_k \leq w(kT) \leq \bar w_k) \approx 1 - \alpha = \beta,
\end{equation}
where
\begin{subequations}
	\begin{align}
		\munderbar w_k & \triangleq \textbf{m}_\text{p}(kT) - z_{\alpha/2}\sqrt{\textbf{C}_\text{p}(kT)}, \\
		\bar w_k       & \triangleq \textbf{m}_\text{p}(kT) + z_{\alpha/2}\sqrt{\textbf{C}_\text{p}(kT)},
	\end{align}
\end{subequations}
and $z_{\alpha/2}$ is the critical value and is obtained by
\begin{equation}\label{eq:risk}
	z_{\alpha/2} \triangleq \Phi^{-1} \big( 1 - \frac{\alpha}{2} \big).
\end{equation}
Here, $\Phi^{-1}$ is the inverse cumulative distribution function of the Gaussian distribution. Critical values for different error rates and confidence levels are listed in Table~\ref{table:CV}.
\begin{table}
	\centering
	\begin{tabular}{c c c}
		\hline
		Confidence Level ($\beta$) & Error Rate ($\alpha$) & Critical value $z_{\alpha/2}$ \\ [0.5ex]
		\hline
		90\%                       & 0.10                  & 1.645                         \\
		95\%                       & 0.05                  & 1.96                          \\
		99\%                       & 0.01                  & 2.575                         \\
		\hline
	\end{tabular}
	\caption{Critical values for commonly used confidence levels \cite{LeBlanc2004}} \label{table:CV}
\end{table}
Now, it is possible to define an envelope for the future disturbances using a GP model given a confidence level:
\begin{equation} \label{eq:envelope}
	\mathbb{W}_k = \mathcal{GP}(\textbf{m}_\text{p}(\textbf{t}_\text{p}), \textbf{C}_\text{p}(\textbf{t}_\text{p}))|\beta.
\end{equation}
In other words, the problem of finding an optimal envelope reduces to acquiring a proper Gaussian process model.

\subsubsection{Selection of hyperparameters}
For improving the prediction accuracy and finding the optimal values for hyperparameters, training of hyperparameters is carried out by maximizing the logarithmic marginal likelihood of training set as

\begin{align}
	\hbar^* & \triangleq \arg \underset{\hbar}{\max} \log p(w(\mathbf{t})|\hbar), \label{eq:trainingobj}
\end{align}
where $\hbar^*$ is the hyperparameters which maximize the log-likelihood of the prediction defined as follows (see \cite{Rasmussen2006} for details):
\begin{align}
	\log p(w(\mathbf{t})|\hbar) & \triangleq -\frac{1}{2} w(\mathbf{t})^T (\textbf{K}_\hbar + \sigma^2I)^{-1} w(\mathbf{t}) - \frac{1}{2} \log \det(\textbf{K}_\hbar + \sigma^2I) - \text{const.}~. \label{eq:training}
\end{align}
Here, $\textbf{K}_\hbar$ is the covariance matrix obtained from evaluating the kernel pairwise at all points of the training set.\\

{\color{black}
		Thus far, for incorporating the forgetting factor concept into Gaussian process models and dealing with non-stationary signals, several techniques have been employed \cite{Perez-Cruz2013}, all of which involve updating the covariance matrix after getting new data.

		A different approach is proposed here for time series data. In order to reduce the impact of old data on prediction, it is assumed that the value of the measured data lose their influence on future as time passes.	Therefore, at the training step to obtain optimal hyperparameters, the likelihood is modified as follows to amplify the uncertainty of the old data points so that they can contribute less to the value of the logarithmic marginal likelihood in (\ref{eq:trainingobj}).
		%for adding the forgetting factor to the model, we need to consider that the likelihood of the data change as we get farther from it. so we multiply the likelihood by the below expression
		\begin{align}
			p(w(\mathbf{t})|\mathbf{w}_k, \hbar)_{ff} & \triangleq p(w(\mathbf{t})|\mathbf{w}_k, \hbar) \times \mathcal{N}(w(\mathbf{t})|\mathbf{w}_k, \textbf{D}), \label{eq:ff}
		\end{align}
		where 
		\begin{align}
		\textbf{D}=\mathrm{diag}(\kappa d_1^\lambda, \kappa d_2^\lambda, \cdots, \kappa d_{N_t}^\lambda), \label{eq:ff2}
		\end{align}
		and $d_i = (t_i - t_{now})$ is the difference between the data point's measured time and the current moment. It is also possible to specify the decay rate by tuning $\kappa \in \mathbb{R}_{+} $ and $\lambda \in \mathbb{R}_{+}$ parameters.
		The new likelihood is also of a Gaussian distribution due to our choice of the multiplier term. Hence, equation~(\ref{eq:training}) can be updated easily to reflect the forgetting factor term as follows:
		\begin{align}
			\log p(w(\mathbf{t})|\hbar)_{ff} & = -\frac{1}{2} w(\mathbf{t})^T (\textbf{K}_\hbar + \sigma^2I+ \textbf{D})^{-1} w(\mathbf{t}) \nonumber            \\
			                        & - \frac{1}{2} \log \det(\textbf{K}_\hbar + \sigma^2I+ \textbf{D}) - \text{const.}~. \label{eq:training2}
		\end{align}
}

Since the marginal likelihoods in~(\ref{eq:training}) and (\ref{eq:training2}) are not convex with respect to $\hbar$, it may have multiple local minima. Thus, the more hyperparameters there are, the less likely it is to successfully find the global minimum with conventional local optimization solvers.

% \subsection{Nonlinear Auto-Regressive method}
% \subsection{Kernel Composition method}
% % Section 3.1: 
% % Section 3.2: NAR method (input is past/present w, output is W_k)
% % Section 3.3: KC method (input is past/present w, output is W_k)
% % Section 3.4: Switching rule 
% and the hyperparameters are saved to be used as the initial point for the training of the next iteration.
% and a new random initial point will be used for training.
\subsection{Review of existing prediction methods}
Gaussian process regression is highly effective for interpolation. Yet, it cannot directly be employed for predicting the future (extrapolation), especially when it comes to the long-term prediction which is needed in some applications for REMPC. Two well-known methods for long-term prediction are the Nonlinear Auto-regressive method and the Kernel Composition method. Each of these methods has its pros and cons. The proposed hybrid method, as shown in Fig.~\ref{fig:diagram2}, combines NAR and KC methods to achieve to the best possible performance. Two mentioned methods and the switching rule are described in the remainder of this section.
\subsubsection{Nonlinear Auto-regressive method}

\begin{figure}
	\centering
	\includegraphics[width=0.5\linewidth]{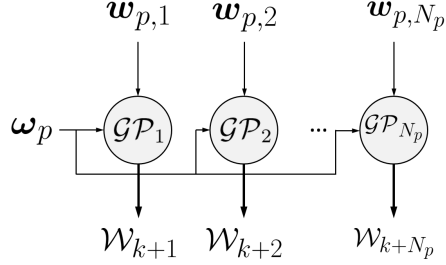}
	\caption{Multiple Gaussian process models in the nonlinear auto-regressive method}
	\label{fig:schematic2}
\end{figure}

In the NAR method, for each time-step ahead, we construct a specific Gaussian process model as shown in Fig.~\ref{fig:schematic2}. It can also be written in a mathematical form as follows:

\begin{equation}
	\omega({t+\chi}) \sim \mathcal{GP}_\chi(\textbf{m}_\text{p}(t+\chi), \textbf{C}_\text{p}(t + \chi)),
\end{equation}
where $\chi \in \left[ 1, 2, \cdots, N_p\right]$ is the prediction time step. In addition, instead of the past time vector $\textbf{t}_\text{p}$ in~(\ref{eq:meancov}), previous values of the disturbances are given to the GP model as independence features. Therefore, input data points to the GP model at time $k$ becomes a matrix in the following form:
\begin{equation}
	\boldsymbol{w}_{p,\chi} =
	\begin{bmatrix}
		w_{k-\chi}     & w_{k-\chi-1}     & \cdots & w_{k-\chi-p+1}     \\
		w_{k-\chi-1}   & w_{k-\chi-2}     & \cdots & w_{k-\chi-p}       \\
		\vdots         & \vdots           & \ddots & \vdots             \\
		w_{k-\chi-N_t} & w_{k-\chi-N_t-1} & \cdots & w_{k-\chi-p-N_t+1}
	\end{bmatrix}
	\in \mathbb{R}^{(N_t+1) \times p}.
\end{equation}
The GP model can capture the dynamics of the disturbance if there is any, and the variable $p$ denotes the order of the model. The corresponding output, instead of $w(\boldsymbol{t}_\text{p})$, for training the model is
\begin{equation}
	\boldsymbol{\omega}_p = \left[ w_{k}, w_{k-1}, \cdots, w_{k-N_t} \right]^{T}.
\end{equation}
The vector $\boldsymbol{w}_{p}$ (dependent variable) contains the last ${N_t+1}^{th}$ data points, and there is a corresponding row in $\boldsymbol{w}_{p,\chi}$ (independent variable), shifted by $\chi$, for each value in $\boldsymbol{w}_{p}$. As a result, the uncertainty propagation through prediction is more reasonable than a Na\"ive multiple-step-ahead prediction. % Y_\chi = [f(t-\chi), \cdots, f(t-\chi-\tau) ] $\tau$ also is our training horizon.

There are several GP models in this approach, as shown in Fig.~\ref{fig:schematic2}, all of which require tuning to have appropriate hyperparameters. Each GP model, however, uses a unique kernel function, such as RBF in (\ref{eq:RBFkernel}). Therefore, this method adds no extra complexity to the GP models and, the training optimization in~(\ref{eq:training}) remains similar to the Na\"ive method. A downside to this method is that there is no connection between these independent models. Thus, the predicted uncertainty for a predicted disturbances might be inconsistent with the neighbors' predictions as demonstrated in Fig.~\ref{fig:NAR}. In this figure, the training horizon is 50 and the prediction horizon is 50. Thus, the NAR model consists of 50 independent GP model, due to the length of prediction horizon, to predict the values of the signal for next 50 steps. The red line is the mean values of the prediction and the red shaded area correspond to the 95\% confidence region.

\subsubsection{Kernel Composition method}
\begin{figure}[h]
	\centering
	\begin{subfigure}[b]{0.7\textwidth}
		\includegraphics[width=\textwidth]{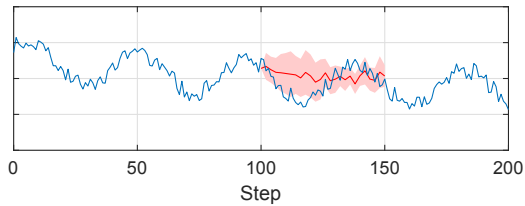}
		\caption{Nonlinear Auto-regressive}
		\label{fig:NAR}
	\end{subfigure}
	%add desired spacing between images, e. g. ~, \quad, \qquad, \hfill etc.
	%(or a blank line to force the subfigure onto a new line)
	\begin{subfigure}[b]{0.7\textwidth}
		\includegraphics[width=\textwidth]{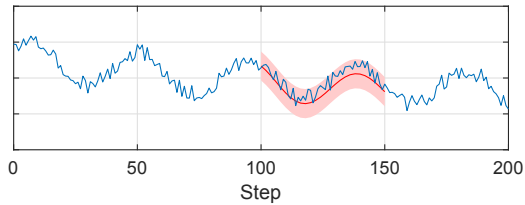}
		\caption{Kernel Composition}
		\label{fig:KC}
	\end{subfigure}
	\caption{Comparison between the NAR and KC methods (the red shaded areas correspond to the 95\% confidence region)}\label{fig:Compare}
\end{figure}

The idea behind the kernel composition method is that each kernel function is responsible for modeling a particular pattern that exists in a time series. Given the extracted patterns, extrapolation for the future and long-term forecast are expected to become precise. As illustrated in Fig.~\ref{fig:KC}, not only the mean values, which is drawn with a red line, follows the signal accurately, the uncertainties of the predicted disturbances are also evaluated more precisely compared to the NAR method.

In general, the composition of the kernel functions can be a complex structure containing summation or multiplication of several kernels. Here, we focus on the additive model which states that a time series can be assembled from trends, seasons, and residual components through addition in the form of the following equation \cite{Duvenaud2011}:

\begin{equation} \label{eq:compoundkernel}
	\textbf{K}(\textbf{t}, \textbf{t}') = \textbf{K}_1(\textbf{t}, \textbf{t}') + \textbf{K}_2(\textbf{t}, \textbf{t}') + \cdots + \textbf{K}_n(\textbf{t}, \textbf{t}'),
\end{equation}
where $n$ is the number of the kernels. Therefore, to obtain the optimal envelope $\mathbb{W}_k$, the kernel in~(\ref{eq:meancov}) needs to be substituted by the compound kernel in~(\ref{eq:compoundkernel}). In addition to the RBF kernel in~(\ref{eq:RBFkernel}), there are quite a few kernel functions used to model patterns. The most common ones are the linear kernel to model linear trends, the periodic kernel to model cyclic behaviors, and a constant kernel to set the mean value of the prediction to a non-zero value. These kernels are calculated as follows:
\begin{align}
	\left[ \textbf{K}_{\hbar=[\zeta]} ^ {\text{Linear}} \right]_{i,j} (\textbf{t}, \textbf{t}')              & \triangleq \frac{\textbf{t}_i \textbf{t}'_j}{\zeta^2},                                               \\
	\left[ \textbf{K}_{\hbar=[\theta,\tau,\psi]} ^ {\text{Periodic}} \right]_{i,j} (\textbf{t}, \textbf{t}') & \triangleq \theta^2  \exp \left( \frac{-2\sin^2( \pi \dfrac{\textbf{t}_i-\textbf{t}'_j}{\tau})}{\psi^2} \right), \\
	\left[ \textbf{K}_{\hbar=[\upsilon]} ^ {\text{Constant}} \right]_{i,j} (\textbf{t}, \textbf{t}')         & \triangleq \upsilon.
\end{align}

%Note that a composed kernel is more complicated compared to the constructing kernels due to the number of hyperparameters.
Note that the composed kernel has more hyperparameters compared to each basic kernel individually. Therefore, the training step becomes a challenge in this method, especially for  real-time applications such as MPC.
\subsection{Switching rule}
\begin{figure}[!h]
	\centering
	\begin{subfigure}[b]{0.6\textwidth}
		\includegraphics[width=\textwidth]{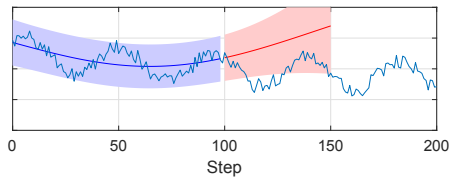}
		\caption{Wrong pattern}
		\label{fig:KCPattern}
	\end{subfigure}
	~ %add desired spacing between images, e. g. ~, \quad, \qquad, \hfill etc.
	%(or a blank line to force the subfigure onto a new line)
	\begin{subfigure}[b]{0.6\textwidth}
		\includegraphics[width=\textwidth]{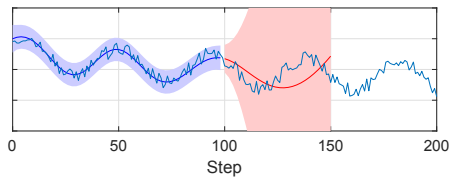}
		\caption{Overfit on training data}
		\label{fig:KCoverfit}
	\end{subfigure}
	\caption{Failure of the KC method (The red shaded areas correspond to the 95\% confidence region)}\label{fig:KCFailure}
\end{figure}

The key issue in designing a predictor based on the hybrid method is the switching rule by which the performance of the overall system is determined. Generally speaking, the KC method fails at the training step due to two main reasons. First, the model misses capturing a correct pattern as shown in Fig.~\ref{fig:KCPattern}. Second, the model overfits training data and cannot predict the future appropriately as shown in Fig.~\ref{fig:KCoverfit}. In the former case, the standard deviation of the training data is different from that of the mean value of the prediction data. In the latter case, as time goes by, the uncertainty increases. This observation can be formulated as two inequalities.

\begin{subequations} \label{eq:switching}
	\begin{align}
		\left| \dfrac{\text{std}(\textbf{w})}{\text{std}(\textbf{m}_\text{p}(\textbf{t}_\text{p}))} -1 \right| & \leq \delta_1^2, \label{eq:switch1} \\
		\left| \dfrac{[\textbf{C}_\text{p}]_{1,1}}{[\textbf{C}_\text{p}]_{N_p,N_p}} -1 \right|                 & \leq \delta_2^2, \label{eq:switch2}
	\end{align}
\end{subequations}
where $\textbf{w}$, $\delta_1$, and $\delta_2$ are the past measured disturbances and tuning parameters respectively. The function $\text{std}(x)$ returns the standard deviation of the vector $x$, and $\textbf{m}_\text{f}$ and $\textbf{C}_\text{f}$ were defined in~(\ref{eq:meancov}). Based on~(\ref{eq:switching})%\footnote{Note that in the case of overfitting, it is also possible to identify the failed models by	performing the validation test with a portion of the measured data and evaluating the	result. However, it needs to ignore a section of the measured data in the training step, thereby reducing the available information for the training.}
, the switching rule is defined as follows:

\begin{enumerate}
	\item If both (\ref{eq:switch1}) and (\ref{eq:switch2}) hold true, the prediction of the KC method is valid, and the predictor chooses this prediction as the output,
	\item If either of the inequalities does not hold true, the prediction of KC method is not valid, and the predictor switches to the NAR method for predicting disturbances.
\end{enumerate}

The smaller that the values of $\delta_1$ and $\delta_2$ are, the hybrid method tends to use the NAR method more frequently. These values should be chosen such that the switching rule be capable of distinguishing the failed models from proper ones.
% the more conservative the hybrid method becomes; which reduces the usage of the internal KC method.
Despite its excellent performance, the behavior of the KC method is susceptible to poor training. In other words, once the training optimization finds a local minimum rather than a global minimum, the prediction of the KC can degrades drastically. Moreover, the KC model only predicts the disturbance well when the structure of disturbances are uniform through the entire training time span. In other words, the KC model may not be able to deal with sudden changes in disturbances. On the other hand, the NAR method, in spite of its relatively poor estimation of uncertainty, solves a simple optimization problem in the training step due to fewer hyperparameters. Note that this switching rule is not perfect, that is, a prediction may satisfy both switching rule inequalities and does not represent the future disturbance precisely. However, by satisfying the switching rule inequalities, one can be sure that the uncertainty envelope does not have an absurd, meaningless behavior similar to Fig.~\ref{fig:KCPattern}.

%\STATE <text>
%\IF{<condition>} \STATE {<text>} \ELSE \STATE{<text>} \ENDIF
%\IF{<condition>} \STATE {<text>} \ELSIF{<condition>} \STATE{<text>} \ENDIF
%\FOR{<condition>} \STATE {<text>} \ENDFOR
%\FOR{<condition> \TO <condition> } \STATE {<text>} \ENDFOR
%\FORALL{<condition>} \STATE{<text>} \ENDFOR
%\WHILE{<condition>} \STATE{<text>} \ENDWHILE
%\REPEAT \STATE{<text>} \UNTIL{<condition>}
%\LOOP \STATE{<text>} \ENDLOOP
%\REQUIRE <text>
%\ENSURE <text>
%\RETURN <text>
%\PRINT <text>
%\COMMENT{<text>}
%\AND, \OR, \XOR, \NOT, \TO, \TRUE, \FALSE
The developed REMPC can be summarized in Algorithm~\ref{alg1}. {\color{black} Saving a proper set of hyperparameters in the proposed algorithm, when the prediction seems satisfactory, helps the hybrid method to have a consistent valid prediction as long as the disturbance's patterns do not vary.}
\begin{algorithm}[h] % enter the algorithm environment
	\caption{REMPC-GP} % give the algorithm a caption
	\label{alg1} % and a label for \ref{} commands later in the document
	\begin{algorithmic} % enter the algorithmic environment
		%\STATE \textbf{Input:} $\hat{\textbf{x}}$ and $\hat{\textbf{w}}$ \qquad \qquad // new measured states and disturbances
		\STATE \textbf{Input:} New measured states and disturbances, $\hat{\textbf{x}}_k$ and $\hat{\textbf{w}}_k$
		\STATE \textbf{Results:} Decision variables, $u_k$
		\STATE
		\STATE Predict using the KC method: [$\textbf{m}_\text{p}$ ,$\textbf{C}_\text{p}$, $\hbar$] = KC($\hat{\textbf{w}_k}$, $\hbar_{in}$)
		\IF{(\ref{eq:switch1})==\TRUE~\AND (\ref{eq:switch2})==\TRUE}
		\STATE Save hyperparameters: $\hbar_{in} = \hbar$
		\ELSE
		\STATE Predict using the NAR method: [$\textbf{m}_\text{p}$ , $\textbf{C}_\text{p}$] = NAR($\hat{\textbf{w}}_k$)
		\STATE Generate random initial point for $\hbar_{in}$
		\ENDIF
		%\STATE $\textbf{w}_{worst}$ = WC($\munderbar{\textbf{w}}$, $\bar{\textbf{w}}$)  \qquad \qquad // find the worst case disturbance scenario
		\STATE $\mathbb{W}_k = \mathcal{GP}(\textbf{m}_\text{p}(\textbf{t}_\text{p}), \textbf{C}_\text{p}(\textbf{t}_\text{p}))|\beta$
		\STATE $u_k = \text{REMPC}(\hat{\textbf{x}}_k, \mathbb{W}_k)$
		%		\ENSURE $y = x^n$
		%		\STATE $y \Leftarrow 1$
		%		\IF{$n < 0$}
		%		\STATE $X \Leftarrow 1 / x$
		%		\STATE $N \Leftarrow -n$
		%		\ELSE
		%		\STATE $X \Leftarrow x$
		%		\STATE $N \Leftarrow n$
		%		\ENDIF
		%		\WHILE{$N \neq 0$}
		%		\IF{$N$ is even}
		%		\STATE $X \Leftarrow X \times X$
		%		\STATE $N \Leftarrow N / 2$
		%		\ELSE[$N$ is odd]
		%		\STATE $y \Leftarrow y \times X$
		%		\STATE $N \Leftarrow N - 1$
		%		\ENDIF
		%		\ENDWHILE
	\end{algorithmic}
\end{algorithm}

\section{Case study} \label{sec:case}
{\color{black} In this section, we apply the proposed EMPC method using disturbance prediction based on hybrid GP to a simple control problem considering various scenarios for the disturbance in order to illustrate its performance.}
\subsection{System description}
\begin{figure}[h]
	\centering
	\includegraphics[width=0.4\linewidth]{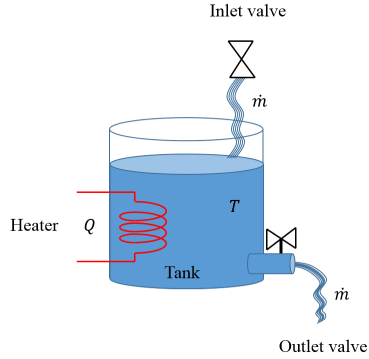}
	\caption{Schematic diagram of the tank-heater system}
	\label{fig:schematic}
\end{figure}

The system considered here consists of a water tank, a heater, and an inlet valve as shown in Fig.~\ref{fig:schematic}. A heater with the variable output power of $Q$ [\si{\kilo \watt}] is connected to the tank to heat the water inside it. It is assumed that the temperature in the tank is uniform. The control objective for this system is to keep the temperature of the water inside the tank, $T$ [\si{\celsius}], higher than 55\si{\celsius} while consuming electrical energy in heater as less as possible. The water enters the tank at the flow rate $\dot{m}$ [\si{\kilo \gram \per \second}] as a disturbance. This unknown disturbance affects the output temperature if the temperature of the inlet flow rate differs from the temperature of the water inside the tank. The heat loss from the tank surface to the ambient is also taken into account. It is assumed that the outlet flow rate is the same as the inlet flow rate for simplicity.

From the first law of thermodynamics, we have the following nonlinear governing equation:
\begin{equation} \label{eq:govequ}
	\dot{T} = f_c(T, Q, \dot m),
\end{equation}
where
\begin{equation}
	f_c(T, Q, \dot m) := \frac{1}{M c_p}\left[ Q - \dot{m} c_p (T - T_{inlet}) - U_{total}(T - T_{amb})\right], \label{eq:govequ2}
\end{equation}
and the descriptions and the values of all the parameters are listed in Table~\ref{tab:parameters}.
\begin{table}[h]
	\begin{center}
		\begin{tabular}{c c c c}
			\hline
			Parameter   & Description                         & Value  & Unit                                          \\ [0.5ex]
			\hline
			$M$         & Total mass of water inside the tank & 0.7854  & \si{\kilogram}                                \\
			\hline
			$c_p$       & Specific heat capacity of water     & 6.9244 & \si{\kilo \joule \per \kilogram \per \kelvin} \\
			\hline
			$T_{inlet}$ & Temperature of inlet flow      & 20     & \si{\celsius}                                 \\
			\hline
			$T_{amb}$   & Ambient temperature                 & 15     & {\celsius}                                    \\
			\hline
			$U_{total}$ & Overall heat loss coefficient       & $10^{-7}$   & \si{\kilo \watt \per \kelvin}                 \\
			\hline
		\end{tabular}
		\caption{Parameters of the tank-heater system} \label{tab:parameters}
	\end{center}
\end{table}

Given~(\ref{eq:govequ}) and $T_{inlet}$ = 20~\si{\celsius}, it is trivial that the worst-case disturbance scenario occurs at the maximum value of the disturbance as our goal is to maintain the water temperature above 55 \si{\celsius}. Thus, finding the worst-case disturbance scenario reduces to estimating the maximum value for the inlet flow rate $\dot{m}$.
%the objective function, $J$, in this range in the range of a prediction horizon is monotonic with respect to the disturbances,
%\begin{equation}
%\text{sign}(\frac{d}{dw_k}J) = Cte.
%\end{equation}
%the worst case happens on either the lowest value or the highest value of the disturbances and
\subsection{Simulation settings}
\begin{figure}
	\centering
	\begin{subfigure}[b]{0.45\textwidth}
		\includegraphics[width=\textwidth]{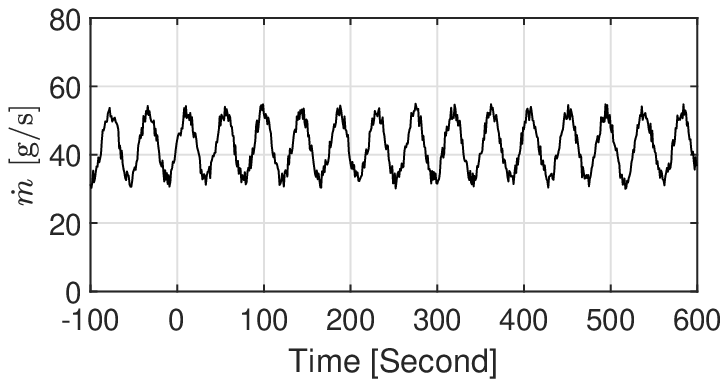}
		\caption{\textbf{SN}}
		\label{fig:sin}
	\end{subfigure}
	~ %add desired spacing between images, e. g. ~, \quad, \qquad, \hfill etc.
	%(or a blank line to force the subfigure onto a new line)
	\begin{subfigure}[b]{0.45\textwidth}
		\includegraphics[width=\textwidth]{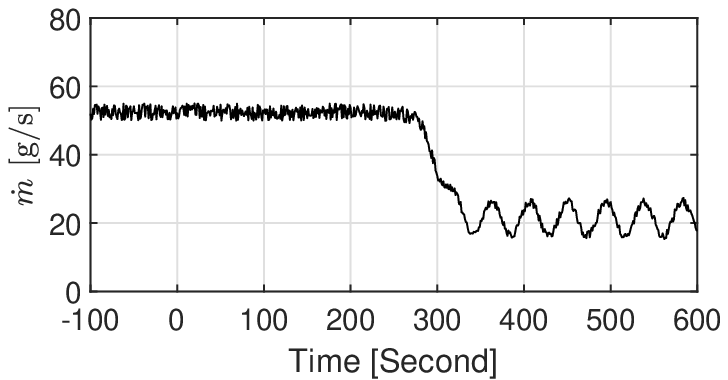}
		\caption{\textbf{LS}}
		\label{fig:linsin}
	\end{subfigure}
	~ %add desired spacing between images, e. g. ~, \quad, \qquad, \hfill etc.
	%(or a blank line to force the subfigure onto a new line)
	\begin{subfigure}[b]{0.45\textwidth}
		\includegraphics[width=\textwidth]{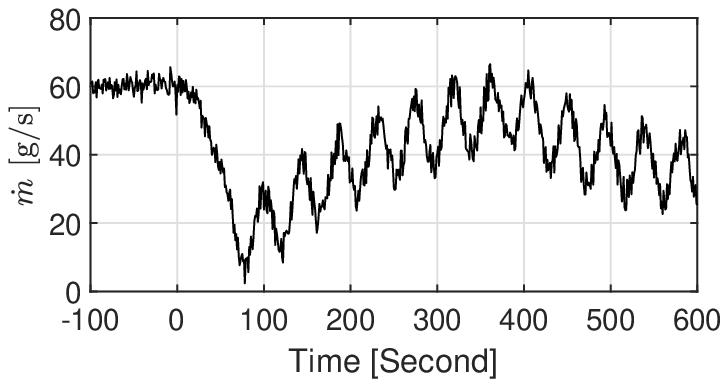}
		\caption{\textbf{CM}}
		\label{fig:com}
	\end{subfigure}
	~ %add desired spacing between images, e. g. ~, \quad, \qquad, \hfill etc.
	%(or a blank line to force the subfigure onto a new line)
	\begin{subfigure}[b]{0.45\textwidth}
		\includegraphics[width=\textwidth]{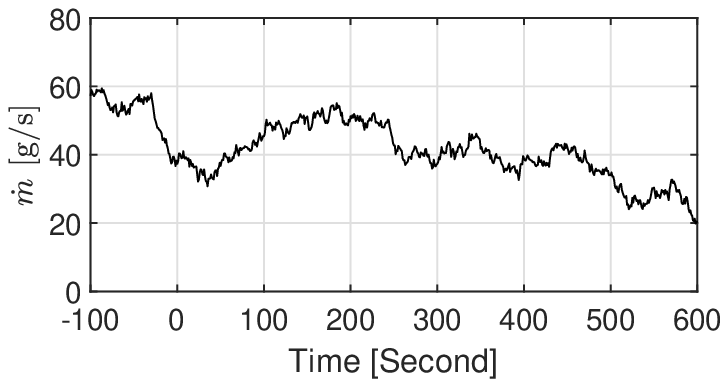}
		\caption{\textbf{RW}}
		\label{fig:rw1}
	\end{subfigure}
	\caption{\color{black} Four different planned scenarios for the inlet mass flow rate, $\dot{m}$}\label{fig:disturbance_scenarios}
\end{figure}

{\color{black} Four different cases are studied here to investigate the performance of the proposed hybrid method for multiple possible disturbance scenarios with different inlet mass flow rate variations.
	\begin{enumerate}
		\item \textbf{SN}\footnote{The bold font words indicate the names of the planned disturbances, to be used in this paper hereafter.}: In this case, inlet mass flow rate is a simple sinusoidal function~(Fig.~\ref{fig:sin}).
		\item \textbf{LS}: At the beginning, the inlet mass flow rate is almost constant. Then, around 300 seconds, its pattern changes to a sinusoidal wave~(Fig.~\ref{fig:linsin}).
		\item \textbf{CM}: In this case, the disturbance is divided into three separate regions. It is considered to be relatively constant from 100 seconds before turning on the controllers until 160 seconds. The disturbance then changes to a periodic flow rate with a varying positive slope on average which continues up to 400 seconds. Within the final region, it remains periodic but with a negative slope on average~(Fig.~\ref{fig:com}).
		\item \textbf{RW}: To assess the performance of the proposed method, when there are no discernible patterns for disturbances, a random walk disturbance flow rate signal is generated~(Fig.~\ref{fig:rw1}).
	\end{enumerate}
	It should mentioned additive noise is added to all disturbances to make prediction more challenging.
}
% A sigmoid function is used to smoothly change the region.

\subsubsection{REMPC description}

The prediction horizon is 50 seconds and the time step is 2~seconds. The economic part of the objective function is a quadratic function of the  heater usage:

\begin{equation}
	\mathbb{J}^{\mathrm{EC}}_{i} = Q_i^2,
\end{equation}
while the constraint violations term~(\ref{eq:CV}) is used with $\munderbar\eta_i = 10$, and $\bar\eta_i = 0$. The maximum and minimum powers of the heater are set to 10 \si{\kilo \watt} and 0 \si{\kilo \watt}, respectively. In addition to the parameters specified to design typical MPC, for the proposed REMPC, a training horizon for disturbance prediction is set to 100 seconds in our simulations. A nonlinear MPC has been written and implemented in MATLAB v2018a using \textit{fmincon} function for solving the optimization problem at each iteration. Parameters of the optimization solver, such as the maximum number of iterations and the termination tolerance for the function value, are set to have the computational time for optimization less than the time step. The equation~(\ref{eq:govequ2}) represents a continuous-time system whereas a discrete-time model is required here. Therefore, the fourth-order Runge-Kutta method is used to discretize~(\ref{eq:govequ2}). The discretized model is as follows:
\setcounter{equation}{31}
\begin{equation}
	T_{n+1} = f_d(T_n, Q_n, \dot{m}_n).
\end{equation}
Provided that the step-size for discretization is within the stability region of the Runge-Kutta method, the obtained discretized model is numerically stable.
%The fourth-order Runge-Kutta is used to discretize~(\ref{eq:govequ}).
\subsubsection{Predictors description}
For the tank-heater system in~(\ref{eq:govequ}) and each disturbance flow rate signal in Fig.~\ref{fig:disturbance_scenarios}, the following five robust economic model predictive controllers with difference predictors are designed:
\begin{enumerate}
	\item \textbf{Perfect}\footnote{The bold font words indicate the names of the designed controllers, to be used in simulation figures and the result section.}: In this case, perfect information on the future disturbance is provided to the controller. Since it gives the best possible performance, it is used to measure the quality of  other controllers.
	\item \textbf{Fixed-range}: In order to compare the REMPC with the varying-range disturbances predictor with the REMPC with a conventional fixed range disturbance uncertainty, used in previous works, $\mathcal{W} = \left[ 0, 70\right]$~\si{\gram \per \second} is regarded as a fixed range for the inlet flow rate.
	\item \textbf{KC}: This controller is based on the KC method with the kernelto be the summation of three terms, i.e., a linear term, a periodic term, and a constant term.
	      \begin{equation}
		      \textbf{K} ^ {\text{KC}}(\textbf{t}, \textbf{t}') = \textbf{K}_{\hbar=[\zeta]} ^ {\text{Linear}}(\textbf{t}, \textbf{t}') + \textbf{K}_{\hbar=[\theta,\tau,\psi]} ^ {\text{Periodic}}(\textbf{t}, \textbf{t}') + \textbf{K}_{\hbar=[\upsilon]} ^ {\text{Constant}}(\textbf{t}, \textbf{t}'). \label{eq:sumker}
	      \end{equation}
	      %	In order to observe the best feasible performance, at each iteration, the training step repeats until proper hyperparameters are obtained considering the proposed switching rule.
	      {\color{black}
	\item \textbf{KC$_{ff}$}: In addition to a standard KC controller, it is trained with the forgetting factor by~(\ref{eq:training2}). Both $\kappa$ and $\lambda$ in~\ref{eq:ff2} are set to $1$.
	      }
	\item \textbf{NAR}: This controller is designed based on the NAR method.%The maximum number of iterations for the optimization for the training is set such that the time required for training does not exceed the time step.
	\item \textbf{Hybrid}: This is the implementation of the proposed hybrid disturbance prediction. In this case, $\delta_1$ and $\delta_2$ in~(\ref{eq:switching}) are set to $\sqrt{0.5}$ and $1$, respectively.
	      {\color{black}
	\item \textbf{Hybrid$_{ff}$}: Similar to \textbf{KC$_{ff}$}, the proposed forgetting factor term is considered for hyperparameters training in addition to the hybrid disturbance predictor.
	      }
\end{enumerate}

Creation and training of the GP models have been realized by using the GPML package \cite{Rasmussen2010}.

\subsection{Results}
In this section, the performance of the proposed forgetting factor term, the proposed hybrid method, and the proposed switching signal are assessed and discussed.

\subsubsection{Comparison between the performance of designed controller}
{\color{black}
The system output, i.e.~the temperature of the water inside the tank, is depicted in Fig.~\ref{fig:result} for four different disturbance profiles. These results, at first glance, demonstrate that the standard KC method is not  able to deal with two distinct patterns in the training horizon; see for example large spikes after 300 seconds in Fig.~\ref{fig:Tlinsin}. Its best performance can be observed in the first case (Fig.~\ref{fig:Tsin}), in which there is no sudden alteration in the disturbance pattern. In other scenarios, whenever the disturbance pattern changes, high peaks appear. As mentioned earlier, the transient behavior also depends on the length of the training horizon in the standard KC method. For instance, a sudden pattern change occurs around 300 seconds in the second case, as seen in Fig.~\ref{fig:linsin}, but the impact of this change on the output temperature lasts up to 400 seconds, as shown in Fig.~\ref{fig:Tlinsin}. Interestingly, the KC method with the forgetting factor, \textbf{KC$_{ff}$}, manages to provide a reasonable prediction even in the presence of different patterns throughout the training horizon. This observation shows the impact of the forgetting factor term on the performance of the KC method.

Having said that, both KC methods are prone to choosing an inappropriate set of hyperparameters. One can notice several small peaks in Fig.~\ref{fig:Tsin} for the KC methods, with and without the forgetting factor, whereas the hybrid methods manages the system such that there is no peaks as high as other methods.

In order to have a better understanding of the performance of each controller, the average value of the objective function in (\ref{eq:obj}) for the entire simulation time is calculated and depicted in Fig.~\ref{fig:compare}.
\begin{figure}
	\centering
	\begin{subfigure}[b]{0.45\textwidth}
		\includegraphics[width=\textwidth]{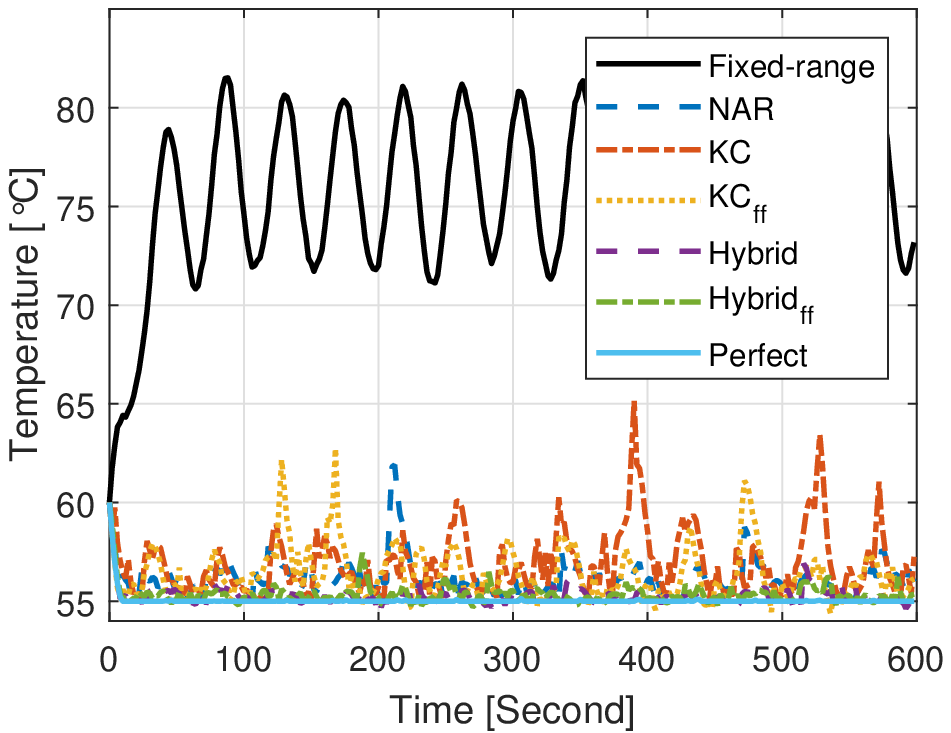}
		\caption{\textbf{SN}}
		\label{fig:Tsin}
	\end{subfigure}
	~ %add desired spacing between images, e. g. ~, \quad, \qquad, \hfill etc.
	%(or a blank line to force the subfigure onto a new line)
	\begin{subfigure}[b]{0.45\textwidth}
		\includegraphics[width=\textwidth]{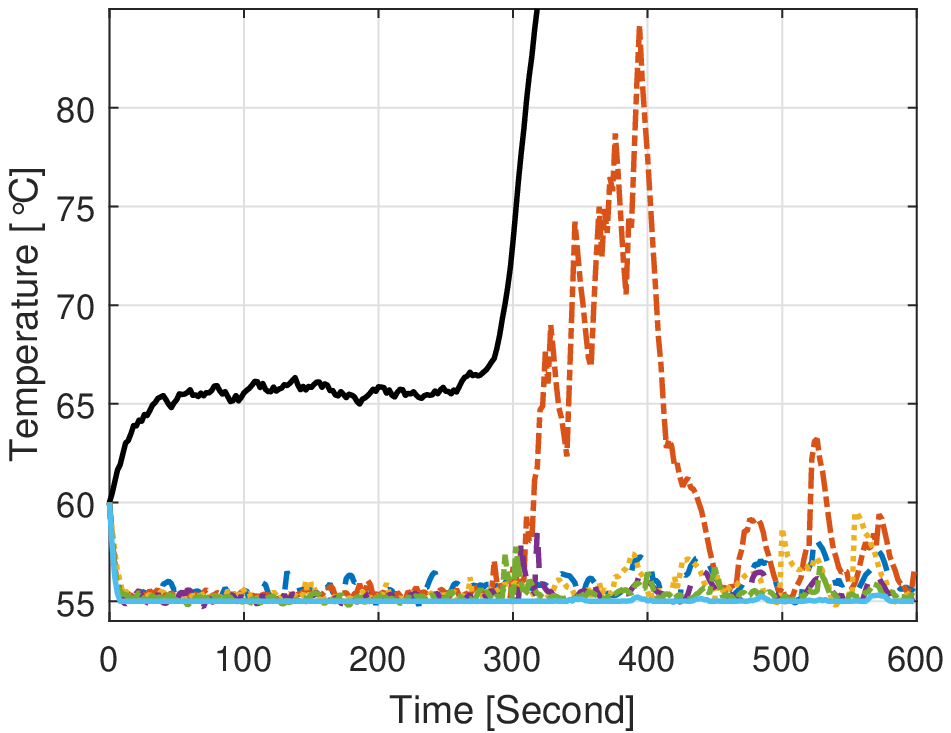}
		\caption{\textbf{LS}}
		\label{fig:Tlinsin}
	\end{subfigure}
	~ %add desired spacing between images, e. g. ~, \quad, \qquad, \hfill etc.
	%(or a blank line to force the subfigure onto a new line)
	\begin{subfigure}[b]{0.45\textwidth}
		\includegraphics[width=\textwidth]{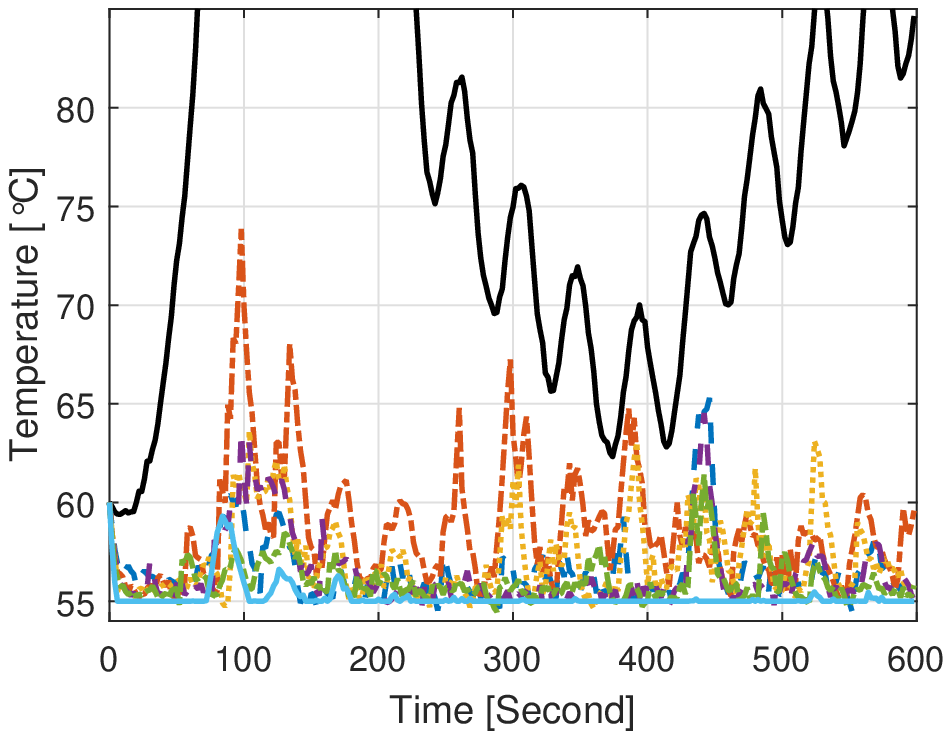}
		\caption{\textbf{CM}}
		\label{fig:Tcom}
	\end{subfigure}
	~ %add desired spacing between images, e. g. ~, \quad, \qquad, \hfill etc.
	%(or a blank line to force the subfigure onto a new line)
	\begin{subfigure}[b]{0.45\textwidth}
		\includegraphics[width=\textwidth]{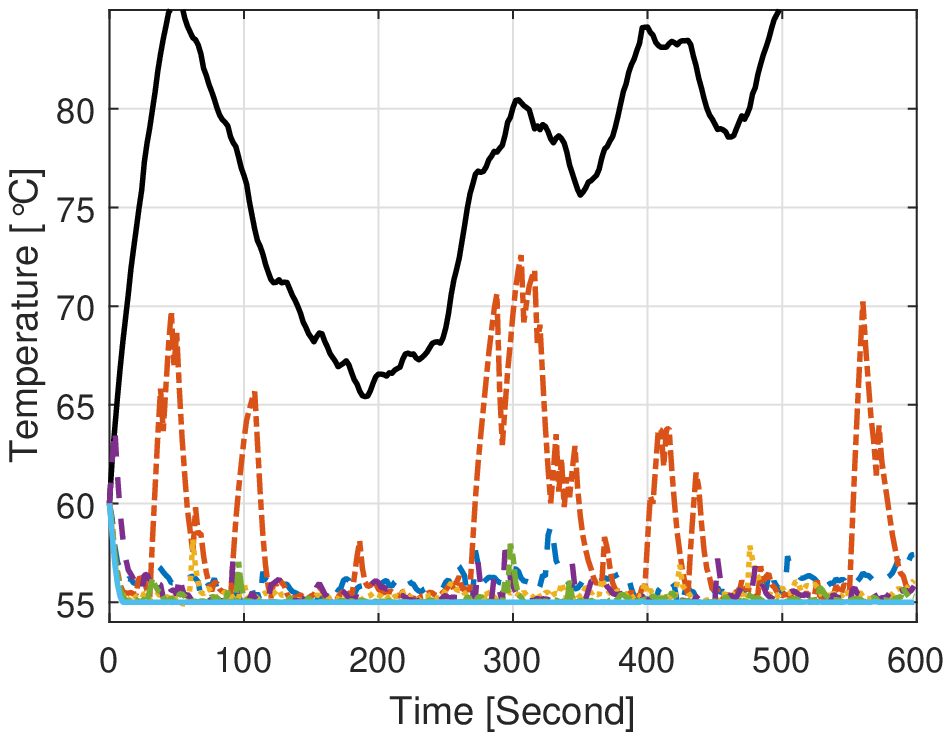}
		\caption{\textbf{RW}}
		\label{fig:Trw1}
	\end{subfigure}
	\caption{\color{black} Temperature of the tank for seven different controllers dealing with four different disturbances}\label{fig:result}
\end{figure}
In the first and second disturbance scenarios, the performances of both hybrid methods are considerably similar to the perfect case. %Also, by adding the forgetting factor term into the likelihood function, the hybrid method's performance slightly degrades as there are only simple patterns in the disturbance, and considering old data with small uncertainty also helps the predictor to improve prediction.
When the disturbance becomes more complicated in the two last scenarios, the difference between the hybrid methods and the perfect case becomes larger. Nonetheless, the hybrid methods still have the best performance among the designed controllers. 
\begin{figure}[h]
	\centering
	\includegraphics[width=0.7\linewidth]{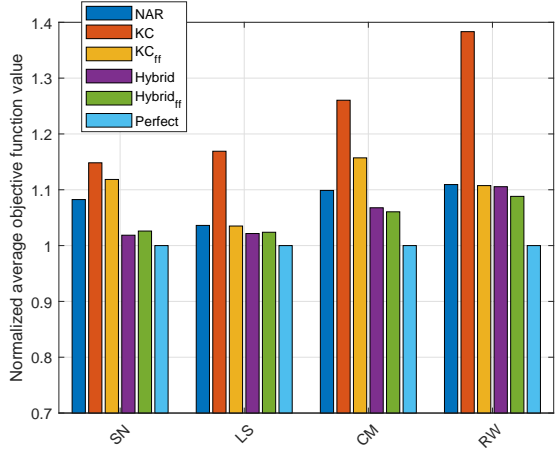}
	\caption{Performance comparison for six designed controllers}
	\label{fig:compare}
\end{figure}

In addition, the amount of thermal power generated by the heater for the \textbf{CM} case is plotted in Fig.~\ref{fig:input}. The fixed-range controller always uses the maximum power of the heater due to its poor foresight. The KC methods also becomes saturated due to the false prediction, yet the hybrid method with the forgetting factor manages the heater input better as there are only a few peaks in the power usage.
\begin{figure}[h]
	\centering
	\includegraphics[width=0.7\linewidth]{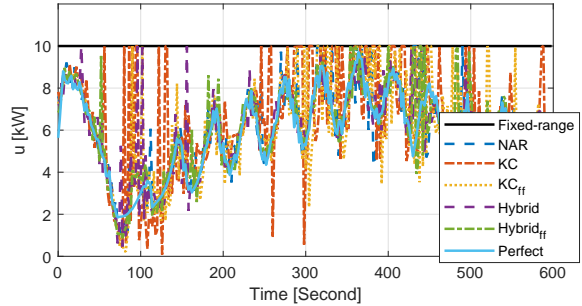}
	\caption{The usage of the heater for six designed controllers in the \textbf{CM} case}
	\label{fig:input}
\end{figure}
%The results show that not only does the hybrid method achieve the closest performance to the perfect controller in terms of constrains violations, it also manages to use less electrical energy compared to the KC and NAR methods. It also shows numerically how many percentages each term increases in comparison to the perfect case.  The values in the last column show the maximum elapsed time required for a single iteration, for both calculating the disturbance range and control input. Despite the fact that the maximum elapsed time for the proposed hybrid method is less than the time step (2 seconds), and thus feasible in this example, this table show that the hybrid estimator has the increased computational load compared to other methods. Therefore, the proposed hybrid estimator is only suitable for relatively slow systems.
\subsubsection{The effect of the length of the training horizon}

The advantage of the hybrid method with forgetting factor over the hybrid method without turns apparent when it comes to the length of the training horizon. As shown in Fig.~\ref{fig:nt}, the training horizon is increased by a factor of 2, 3, and 4 for the \textbf{CM} disturbance profile. The normalized average objective function for the standard hybrid method (without forgetting factor) indicates that there is an optimal length for the training horizon, and after that the performance diminishes since the predictor cannot adapt to recent disturbance patterns. This is not the case for the hybrid method with the forgetting factor as old data systematically lose their weights in the prediction. Therefore, the only limitation for the length of the training horizon comes from the amount of demanded memory and computational time.
\begin{figure}[h]
	\centering
	\includegraphics[width=0.6\linewidth]{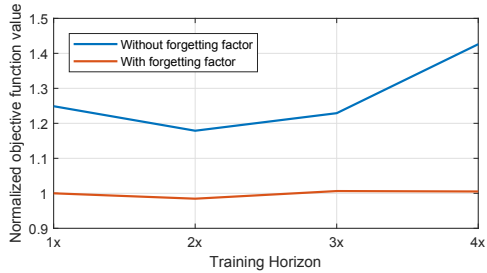}
	\caption{The effect of adding the forgetting factor term to the training equation}
	\label{fig:nt}
\end{figure}

\subsubsection{Performance of the proposed switching rule}

To investigate the performance of the switching rule in the hybrid methods, the switching signal is plotted in Fig.~\ref{fig:signal}. It is observed that the hybrid controller switches to the NAR method once the disturbance changes abruptly as expected, see for instance the switching signals after 300 seconds in Fig.~11b.. Although the hybrid methods and the KC methods usually use the same approach for predicting in the \textbf{SN} and \textbf{LS} disturbance cases, the results are entirely different as shown in Fig.~\ref{fig:result}. The reason for this difference is that there is no criterion in the KC methods to identify the failed-in-training models. Therefore, the hyperparameters are initialized randomly before the training step at each iteration, leading to numerous inaccurate prediction models for prediction. On the other hand, the switching rule in the hybrid methods gives the KC method the opportunity of finding suitable hyperparameters and saves them for the training step at the next iteration.

Despite its satisfactory performance, the switching rule is quite simple in this paper, and as shown in Fig.~\ref{fig:Scom}, at some points, there is successive switching between the two approaches which may deteriorate the control system performances. Therefore, further investigation for a better switching rule is needed.
\begin{figure}[h]
	\centering
	\begin{subfigure}[b]{0.4\textwidth}
		\includegraphics[width=\textwidth]{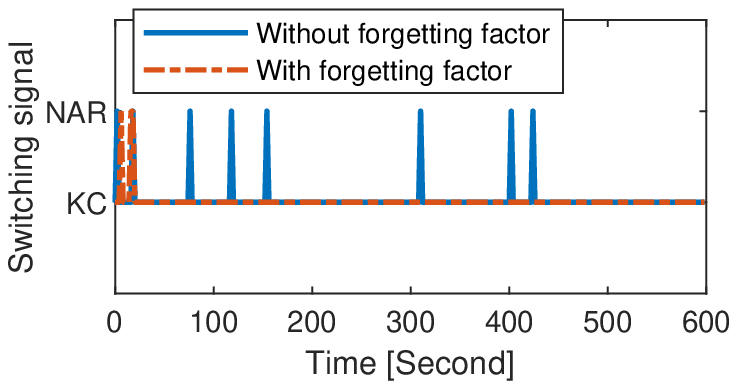}
		\caption{\textbf{SN}}
		\label{fig:Ssin}
	\end{subfigure}
	~ %add desired spacing between images, e. g. ~, \quad, \qquad, \hfill etc.
	%(or a blank line to force the subfigure onto a new line)
	\begin{subfigure}[b]{0.4\textwidth}
		\includegraphics[width=\textwidth]{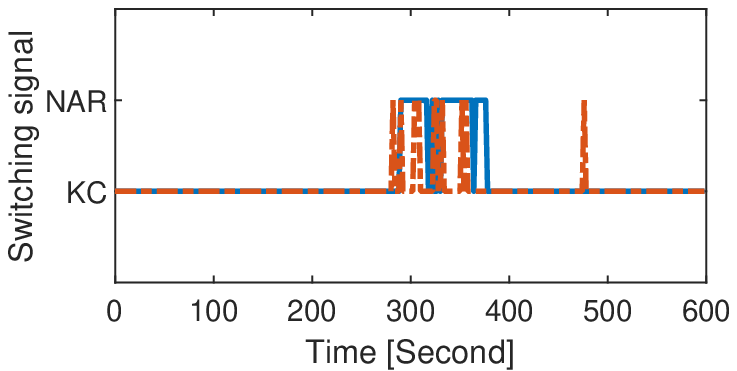}
		\caption{\textbf{LS}}
		\label{fig:Slinsin}
	\end{subfigure}
	~ %add desired spacing between images, e. g. ~, \quad, \qquad, \hfill etc.
	%(or a blank line to force the subfigure onto a new line)
	\begin{subfigure}[b]{0.4\textwidth}
		\includegraphics[width=\textwidth]{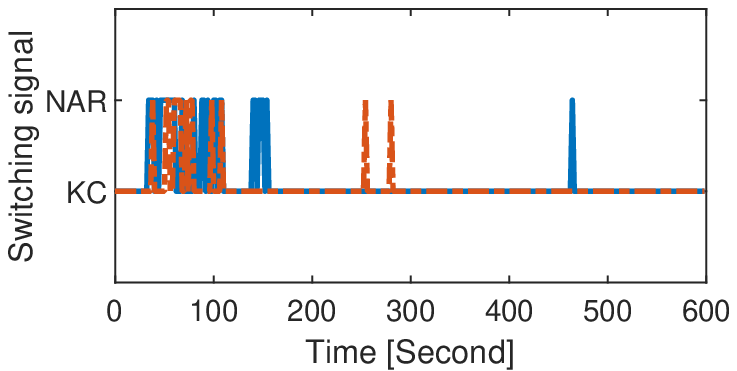}
		\caption{\textbf{CM}}
		\label{fig:Scom}
	\end{subfigure}
	~ %add desired spacing between images, e. g. ~, \quad, \qquad, \hfill etc.
	%(or a blank line to force the subfigure onto a new line)
	\begin{subfigure}[b]{0.4\textwidth}
		\includegraphics[width=\textwidth]{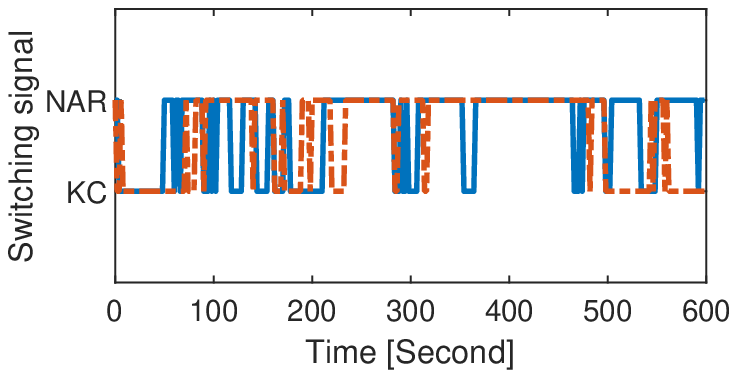}
		\caption{\textbf{RW}}
		\label{fig:Srw1}
	\end{subfigure}
	\caption{\color{black} Switching signal in the hybrid method for four different inlet mass flow rate scenarios, $\dot{m}$}\label{fig:signal}
\end{figure}
\subsubsection{Discussion on the required data storage}

Finally, it should also be mentioned that data storage might be a concern for the proposed predictor due to non-parametric Gaussian process models. The dimensions of the kernel matrix in (\ref{eq:Joint}) are $(N_p+N_t) \times (N_p+N_t)$, and thus we need to store $(N_p+N_t)^2$ values. In the NAR method, since there is a unique model for each step-ahead prediction, the number of the stored values increases to $(N_p+N_t)^2 \times N_p$, and in the hybrid method, it rises to $(N_p+N_t)^2 \times (N_p+1)$. If it is assumed that the order of $N_p$ and $N_t$ are the same, the data storage for the proposed method is of $\mathcal{O}(N_p^3)$. It means the required data storage grows exponentially as the prediction horizon gets longer.

\section{Conclusions}
This paper presented a novel hybrid approach based on the Gaussian process method to address the issue of unknown future disturbances in robust economic model predictive controllers. It also proposed a new approach to incorporate the forgetting factor concept in Gaussian process models. The proposed approach utilized the kernel composition method alongside the nonlinear auto-regressive method. The kernel composition method offers great potential for predicting unknown disturbances based on their inherent patterns, yet it is not suitable for real-time applications. On the other hand, the nonlinear auto-regressive method is more stable, especially when facing sudden changes in patterns, but its prediction is not as precise as the kernel composition method in general. Thus, a switching strategy was employed in the hybrid method to identify inappropriate predictions of the kernel composition method and turn to the nonlinear auto-regressive model. The forgetting factor for the training improves the performance of the obtained Gaussian process model and makes it robust against multi-pattern disturbances. The efficiency of the hybrid method with the forgetting factor and the performance of the switching strategy were validated using a simple water temperature control problem with a heater. In this control problem, the proposed hybrid method achieved the best performance for model predictive controller not only when the disturbance has an obvious pattern, but for a random walk disturbance profile as well. %Possible future research topics are to analyze the transient behavior and to guarantee the stability for such a switching system. 
}
\section*{Acknowledgments}
This work was supported by the Natural Sciences and Engineering Research Council of Canada through the Collaborative Research and Development Grant, Ascent Systems Technologies, and the Four-Year-Fellowship of the University of British Columbia.
\bibliography{mybibfile}

\end{document}